\documentclass[%
 reprint,
 amsmath,amssymb,
 aps,
]{revtex4-1}

\usepackage{graphicx}
\usepackage{dcolumn}
\usepackage{bm}

\newtheorem{prop}{Proposition}
\newtheorem{defi}{Definition}

\newcommand{\munite}{\mathbf{1}}
\newcommand{\ov}[1]{\overline{#1}}
\newcommand{\abs}[1]{|#1|}
\newcommand{\tridente}{\mathrel{\reflectbox{\rotatebox[origin=c]{180}{$\pitchfork$}}}}

\begin{document}


\title{Alternative formulation for the operator algebra over the space of paths in a ADE $SU(3)$ graph.}
\author{Jes\'us A. Pineda} \email{jesuspineda@usb.ve}
\author{Esteban Isasi} \email{eisasi@usb.ve}
\author{Mario Caicedo} \email{micaicedo@usb.ve}


\affiliation{%
 Departamento de F\'isica\\Universidad Sim\'on Bol\'\i var.
}%


\date{\today}

\begin{abstract}
In this work we discuss the elements required for the construction of the operator algebra for the space of paths over a simply laced $SU(3)$ graph. These operators are an important step in the construction of the bialgebra required to find the partition functions of some modular invariant CFTs. We define the cup and cap operators associated with back-and-forth sequences and add them to the creation and annihilation operators in the operator algebra as they are required for the calculation of the full space of essential paths prescribed by the fusion algebra. These operators require collapsed triangular cells that had not been found in previous works; here we provide explicit values for these cells and show their importance in order for the cell system to fulfill the Kuperberg relations for $SU(3)$ tangles. We also find that demanding that our operators satisfy the Temperley-Lieb algebra leads one naturally to consider operators that create and annihilate closed triangular sequences, which in turn provides an alternative the cup and cap operators as they allow one to replace back-and-forth sequences with closed triangular ones. We finally show that the essential paths obtained by using closed triangles are equivalent to those obtained originally using back-and-forth sequences.

\end{abstract}

\maketitle


\section{\label{sec:level1}Introduction}

Over the last two decades or so, research performed in a number of fields in the thin border between theoretical physics and mathematics, suggest that there exists a fundamental structure that constitutes spinal cord of sorts that links topics as diverse as: statistical mechanics, string theory, quantum gravity, conformal field theory, re-normalization in quantum field theory, theory of bimodules, Von Neumann algebras, sector theory, (weak) Hopf algebras, modular categories, etc. to only mention a few.

The corresponding algebraic structures have been analyzed more or less independently by many groups within these fields, with their own tools and terminology. The results obtained by these different research programs are not always easy to compare, or even to comprehend, because of the intricacies of the required backgrounds and specificity of the language of each subdiscipline. However, at the heart of any such description we can always find fundamental objects that can be described as a pair of categories $(\mathcal{A}(G),\mathcal{E}(G))$ associated to a compact Lie group $G$, one of which is a fusion $\mathcal{A}(G)$ category,  while the other is a modular category $\mathcal{E}(G)$. Or, conversely, it can be described as  pair of generalized $ADE$ graphs $(\mathcal{A}(G),\mathcal{E}(G))$.

In one of these fields, conformal field theory, the relation is established via the well known ADE classification of the modular invariant partition functions of $su(N)$ WZW Conformal Field Theories (CFT). The formal structure that lies behind this classification has been developed over time and can be studied thoroughly in \cite{Ocneanu:paths}, \cite{Ocneanu:2000}, \cite{CoqueGarciaTrinchero:1999},\cite{CoqueTrincheroDTE:2004}, \cite{EvansBockenhauerIII:1998}, \cite{EvansPintoSubfqctor:2003}, \cite{Coque6jsymbols:2006}.

The physical-mathematical model that establishes the formal relation between a consistent pair of categories $(\mathcal{A}(G),\mathcal{E}(G))$ (composed of a fusion category and a modular category) with a $SU(3)$ WZW theory, relies on the set of equations that ensures the coherent composition between the maps of the fusion category and the maps of the module category. These consistency equations, usually know as the Small Pocket and Big Pocket equations, were originally proposed by Adrian Ocneanu and have been published and used in a series of examples in \cite{CoquereauxIsasiSchieber:2010}.

The consistency equations, are obtained by embedding two of the three Kuperberg relations (the two that are not associated with the trivial representation) within the intertwines of the module category. The embedding leads to a pair of equations on the values of the ,so-called, triangular cells $T_{abc}^{\alpha \beta \gamma}$ of the graph representing the corresponding module category. The values of the triangular cells for all graphs (modular category) of $SU(3)$ were reported in \cite{CoquereauxIsasiSchieber:2010} and \cite{Evans:2009ar}.

The study of the properties of the cell system of $SU(3)$ graphs and the explicit calculation of the triangular cells that form it, permitted the explicit calculation of the full set of essential paths over a $SU(3)$ graph. In \cite{CaicedoIsasiPineda:2015}, an intuitive geometrical interpretation of the operators involved in the calculation of essential paths over simply laced paths was introduced. This allowed for a clearer understanding of the structure of the space of paths. In that work, many properties of the essential paths were explored and explicitly calculated, which made clear that, in order to properly describe essential paths over a simply laced $SU(3)$ graph it was necessary to include heretofore unknown collapsed triangular cells corresponding to triangles with degenerate edges (or vertices) and a proposal for their values was included therein.

From the perspective of the consistency relations, the need for collapsed cells naturally appears from the Kuperberg relations, in particular from the circle, which is constructed from the intertwiner containing the trivial representation and the generators $\sigma$ and $\ov{\sigma}$ in the $SU(3)$ recoupling model. However, even though the collapsed cells are contained in the consistency relations, collapsed triangles are concealed by the fact that  they are combined in such a way that results in a scalar that is factored away. In order to bring forth clearly the collapsed cells one should ``split'' the circle into the cup and cap operators or in terms of the back-and-forth paths with which they create and annihilate, that explicitly require collapsed triangular cells. This splitting had so far not been considered in the literature in relation to paths on a graph.

In this regard it is important to point out that in order to fully obtain the essential paths over a simply connected $SU(3)$ graph \footnote{We remind the reader that each admissible triangle obtained from the fusion algebra is in one to one correspondence to an essential path.} one requires four operators: The creation and annihilation operators, that are involved in the creation and annihilation of triangular sequences (``open'' triangles, that is sequences of vertices forming two of the three edges of a triangle) and the cup and cap operators that create and annihilate back-and-forth sequences and, as mentioned above, explicitly require collapsed triangular cells. 

In the aforementioned construction there are two types of sequences that are taken to be the fundamental backtracking sequences, these are the open triangular sequence and the back-and-forth sequence. This means that this formalism openly disregards closed triangular sequences as an option for a fundamental backtracking  sequence in a path, even if they would seem natural candidates from the geometrical standpoint. In this way, a closed triangular sequence can be seen as two consecutive open triangular sections and is obtained from a back-and-forth sequence on which then one adds an open triangle. 

In spite of this, closed triangles make an appearance when one demands that the combined operator $U_i=C_iC_i$ fulfill the $SU(3)$ Temperley-Lieb algebra, since the introduction of operators that create and annihilate closed triangles make calculations less involved. In light of this, a natural question then is if one could somehow use closed triangular sequences in the construction of essential paths. In this work we find that there actually are two equivalent formulations for the set of essential paths, one that considers triangular sequences and back-and-forth sequences as the fundamental backtrackings, and another one that exchanges back-and-forth sequences for closed triangular ones. By equivalence here what is meant is that, given an admissible triangle, it is possible to find paths starting and ending in the appropriate vertices and are labeled by an element of the $\mathcal{A}(G)$ graph \footnote{Note that this label can be interpreted as the length of the graph in different ways. This is due to the fact that the path lies in the lattice of the $SU(3)$ graph.} and are expressed in terms of the linear combination of paths with open triangles and back-and-forth sequences or with open and closed triangles.

 

\section{Creation and annihilation of paths}

We begin our discussion by recalling the definition of a path on a graph, this definition has been studied before in the literature \cite{CoquereauxIsasiSchieber:2010,Trinchero:2010yr,Ocneanu:2000}. 

\begin{defi}
\label{defi-path}
An elementary path ($e$) is a sequence $e=v_{0}v_{1}\dots v_{i-1}v_{i}v_{i+1}\dots v_{n}$ of vertices connected by edges which may belong to either one graph or its {\it{conjugate}}. An elementary path can also be defined as a series of consecutive edges (which will be oriented or not depending on the graph) connecting vertices of the graph $e=\xi_{0,1}\xi_{1,2}\dots \xi_{i-1,i}\xi_{i,i+1}\dots \xi_{n-1,n}$. 
\end{defi}

\begin{defi}\label{defi:spacepaths}
The space of paths $\mathcal{P}$ is the inner product space spanned by elementary paths. Elementary paths form an orthonormal basis of this space.
\end{defi}

\begin{defi}\label{defi:elempath}
A path is an element of $\mathcal{P}$, therefore it can be expressed as:

\begin{equation}\label{eq:defielem}
\eta=\sum_{i}\alpha_{i}e_{i}\,,
\end{equation}
where $\{e_i\}$ is the basis of $\mathcal{P}$
\end{defi}

We must call attention to the fact that definition \ref{defi-path} implies that two graphs must be simultaneously considered, indeed, a given sequence of vertices (or edges) can in general contain vertices connected by edges that follow any direction of the arrows (and thus can be understood as elements of the graph $G$ or its conjugate). The length of a path, $l$ will be the number of edges that form the path or the number of vertices minus one, irrespective of the orientation of said edges. 

\subsection{Creation and annihilation operators}

\begin{defi}The path creation and annihilation operators, acting on an elementary path $\eta$ are given by: 
\begin{equation}\label{eq:defaniq}
C_{i}(\eta)= \frac{T_{i-1,i,i+1}}{\sqrt{[i-1][i+1]}}v_{0}v_{1}\dots v_{i-1}v_{i+1}\dots v_{n},
\end{equation}

\begin{equation}\label{eq:defcrea}
C^{\dag}_{i}(\eta)= \sum_{b\,n.n.\,v_i}\frac{\overline{T_{i-1,b,i}}}{\sqrt{[i-1][i]}}v_{0}v_{1}\dots v_{i-1}v_{b}v_{i}\dots v_{n}.
\end{equation}
\end{defi}

These operators add or remove ``open'' triangular sequences of vertices in such a way that the vertex that is added or removed is the middle one of the triangle constructed with the vertices $i-1,i,i+1$. If such a sequence does not exist or if the operator acts on a path of length $L<2$, the annihilation operator $C_{i}$ gives zero result. For the creation operator the sum is over all vertices $v_{b}$ that are common neighbors of $i-1$ and $i$. Here we see that the meaning of a backtracking sequence for $SU(2)$ changes to a triangular sequence in $SU(3)$, and where the prefactor $T_{i-1,i,i+1}$ is the triangular cell associated to elementary triangle $v_{i-1}v_{i}v_{i+1}$ of the graph. The values of these coefficients  for elementary triangles  have been calculated in \cite{Evans:2009ar, CoquereauxIsasiSchieber:2010}

$C_{i}$ and $C^{\dag}_{i}$ are insufficient to construct the complete set of essential paths over the graph (that we need to calculate the bialgebra) \cite{CaicedoIsasiPineda:2015}. This problem can be seen when one considers the possibility of back-and-forth paths analogous to those in $SU(2)$, these paths are not created or annihilated by the above definitions and thus require us to introduce two operators related $C_{i}$ and $C^{\dag}_{i}$, these are the so called cup and cap operators \cite{CaicedoIsasiPineda:2015}.
\begin{defi}The cup $\cup_{i}$ and cap $\cap_{i}$ operators, acting on a given elementary path $\eta$ are:
\begin{equation}\label{eq:defcup}
\cup_{i}(\eta)= \frac{T_{i-1,i,i+1}}{\sqrt{[i-1][i+1]}}\delta_{i-1,i+1}\;v_{0}v_{1}\dots v_{i-1}v_{i+2}\dots v_{n},
\end{equation}
\begin{equation}\label{eq:defcap}
\cap_{i}(\eta)= \sum_{v_b \,n.n.\,v_i}\frac{\overline{T_{i-1,b,i-1}}}{\sqrt{[i-1][i-1]}}v_{0}v_{1}\dots v_{i-1}v_{b}v_{i-1}v_{i}\dots v_{n}.
\end{equation}
\end{defi}
These operators create or annihilate back-and-forth sequences that are not clearly triangular but are in direct analogy to the back-and forth sequences that are treated by the creation and annihilation operators for paths in $SU(2)$ \cite{Trinchero:2010yr}. Whenever the input path lacks one such back-and-forth sequences the action of the cup operator on it yields zero. The sum in the cap operator adds all possible back-and-forth sequences to the nearest neighbors of the $i-1$-th vertex. These operators make use of the so called collapsed triangular cells, that were introduced in \cite{CaicedoIsasiPineda:2015} and have not been studied thoroughly in the literature. 

\begin{defi}
A collapsed (or deformed) triangular sequence  is  a back-and-forth sequence of three vertices $v_{i-1},v_{i},v_{i+1}$ such that  the first an last vertices ($v_{i-1}$ and $v_{i+1}$) are the same.
\end{defi}

\begin{prop}
Given a collapsed triangular sequence $v_{i-1},v_{i},v_{i+1}$,  ($v_{i-1}=v_{i+1}$), the value of the associated cell $T_{i-1, i, i-1}$ is given by 
\begin{equation}
T_{i-1, i, i-1}=\sqrt{[i-1][i]}.
\end{equation}
\end{prop}

With this definition, the graph's cell system includes collapsed cells naturally in such a way that the Kuperberg identities are satisfied or, in an identical fashion, the so-called ``small pocket'' and ``big pocket'' equations. Equations \ref{eq:kup1}, \ref{eq:kup2}, \ref{cuadratico1} and \ref{cuadratico2}, when taken as a whole, provide the explicit proof of the required properties of the collapsed cells in relation to the Kuperberg relations for rank 2 tangles. See   \cite{CaicedoIsasiPineda:2015} for the fully worked out proofs and for calculations using collapsed cells in the process of finding the essential paths for two $SU(3)$ ADE graphs.$\blacksquare$

For completeness, and since we will  repeatedly refer to them, we recall the properties of the triangular cells for simply connected graphs in $SU(3)$. These relations are well established in the literature and are related to the Kuperberg relations for $A_{2}$ tangles \cite{kuperberg1996, CoquereauxIsasiSchieber:2010}:

\paragraph{``Small pocket" or Type~I equations:}
\begin{equation}\label{eq:smallpock}
\sum_c  \abs{T_{abc}}^2 = \beta [a] [b],
\end{equation}

\paragraph{``Big pocket'' or Type~II equations:} These equations read:
\begin{equation}\label{eq:bigpock}
\sum_k \frac{1}{[k]} T_{ij k} \ov{T_{l j k}} T_{l m k} \ov{T_{im k}} = [i][j][l]\delta_{j,m} +  [i][j][m]\delta_{i,l}.
\end{equation}

For the fully worked out calculations of both the properties of the non-collapsed triangular cells and their values for some $SU(3)$ graphs, we refer readers to \cite{Evans:2009ar, CoquereauxIsasiSchieber:2010}.

Let us now check some properties that will be needed to bolster the main proposition of this work. 

\begin{prop}
For a path $\eta$, one finds that the cup, cap, creation and annihilation operators satisfy the following relations:
 \begin{align}
 C_{i}C^{\dag}_{j}&=C^{\dag}_{j-1}C_{i},\quad \mbox{for}\; i<j-1 \\
 C_{i}C^{\dag}_{j}&=C^{\dag}_{j}C_{i-1},\quad \mbox{for}\; i>j+1 \\
  C_{i}C^{\dag}_{i}&=\beta\munite_{n},\quad \mbox{for}\; i\leq n\label{eq:kup1}\\
  \cup_{i\pm 1}\cap_{i}&=\munite,\quad \mbox{for}\; i\leq n\\
 \cup_{i}\cap_{i}&=\alpha\munite_{n},\quad \mbox{for}\; i\leq n\label{eq:kup2}
 \end{align}
\end{prop}

\paragraph{Application of the creation and annihilation operators:} The first three relations pertain to the behaviour of the creation and annihilation operators in all possible relative positions save for $j=i\pm 1$, which we will study momentarily. The second two, using the cup and cap operators, are related to topological simplifications of braids studied by Kauffman in \cite{kauffman1991knots}. 

For the first relation we find:
\begin{align}
 C_{i}C^{\dag}_{j}\eta
&=\sum_{b}\frac{T_{i-1,i,i+1}}{\sqrt{[i-1][i+1]}}\frac{\ov{T_{j-1,b,j}}}{
\sqrt{[j-1][j]}}\times \nonumber\\ &\times\;v_{0}v_{1}\dots v_{i-1}v_{i+1}\dots v_{j-1}v_{b}v_{j}\dots
v_{n}, \\
 C^{\dag}_{j-1}C_{i}\eta &=\sum_{b}\frac{T_{i-1,i,i+1}}{\sqrt{[i-1][i+1]}}\frac{\ov{T_{j-1,b,j}}}{\sqrt{
[j-1][j]}}\times\nonumber\\ &\times\;v_{0}v_{1}\dots v_{i-1}v_{i+1}\dots v_{j-1}v_{b}v_{j}\dots v_{n}.
 \end{align}

For the second one the proof is also straight forward:
 \begin{align}
 C_{i}C^{\dag}_{j}\eta
&=\sum_{b}\frac{\ov{T_{j-1,b,j}}}{
\sqrt{[j-1][j]}}\frac{T_{i-2,i-1,i}}{\sqrt{[i-2][i]}}\times \nonumber\\&\times\;v_{0}v_{1}\dots v_{j-1}v_{b}v_{j}\dots
v_{i-2}v_{i}v_{i+1}\dots v_{n} \\
 C^{\dag}_{j}C_{i-1}\eta &=\sum_{b}\frac{T_{i-2,i-1,i}}{\sqrt{[i-2][i]}}\frac{\ov{T_{j-1,b,j}}}{\sqrt{[j-1][j]}}\times \nonumber\\ &\times \;v_{0}v_{1}\dots v_{j-1}v_{b}v_{j}\dots v_{i-2}v_{i}\dots v_{n}
 \end{align}

\paragraph{Relationship between the application of the operators and the Kuperberg relations:} When both operators act on the same vertex we find a result analogous to the first of the well known Kuperberg relations for $A_{2}$ tangles:
 \begin{align}
 C_{i}C^{\dag}_{i}\eta
&=\sum_{b}\frac{T_{i-1,b,i}\ov{T_{i-1,b,i}}}{[i-1][i]}\;v_{0}v_{1}\dots
v_{i-1}v_{i}\dots v_{n} \nonumber\\
 &=\sum_{b}\frac{\abs{T_{i-1,b,i}}^{2}}{[i-1][i]}\;v_{0}v_{1}\dots
v_{i-1}v_{i}\dots v_{n} \nonumber\\
 &=\beta \eta
 \end{align}
 The values for the triangular cells and their properties, are given by the big and small pocket equations, which can be found proven fully in \cite{CoquereauxIsasiSchieber:2010}.
 
\paragraph{The identity relation for the cup and cap operators:} The first of the cup cap simplifications is immediate since it amounts to adding and subtracting back-and-forth sequences:
\begin{align}
\cup_{i+1}\cap_{i}(\eta)&=\sum_{b}\frac{\ov{T_{i-1,b,i-1}}}{[i-1]}\frac{T_{b,i-i,i}}{\sqrt{[b][i]}}\delta_{b,i}\times\nonumber\\ &\times\;v_{0}v_{1}\dots
v_{i-1}v_{b}v_{i+1}\dots v_{n} \nonumber\\
&=v_{0}v_{1}\dots v_{i-1}v_{i}v_{i+1}\dots v_{n}\\
\cup_{i-1}\cap_{i}(\eta)&=\sum_{b}\frac{\ov{T_{i-1,b,i-1}}}{[i-1]}\frac{T_{i-2,i-1,b}}{\sqrt{[b][i-2]}}\delta_{b,i-2}\times\nonumber\\&\times\;v_{0}v_{1}\dots
v_{b}v_{i-1}v_{i}\dots v_{n} \nonumber\\
&=v_{0}v_{1}\dots v_{i-1}v_{i}v_{i+1}\dots v_{n}
\end{align}
In the above we have used the definition of the collapsed triangular cells.

\paragraph{The tangle circle relation for paths:} The final relation for the cup and cap operators, that adds and removes a back-and-forth sequence in the same point of the path, yields interesting results:
\begin{align}
\cup_{i+1}\cap_{i}(\eta)&=\sum_{b}\frac{\ov{T_{i-1,b,i-1}}}{[i-1]}\frac{T_{i-1,b,i-1}}{[i-1]}\times\nonumber\\&\times\;v_{0}v_{1}\dots v_{i-1}v_{i}v_{i+1}\dots v_{n} \nonumber\\
&=\sum_{b}\frac{\abs{T_{i-1,b,i-1}}^{2}}{[i-1]^{2}}\;v_{0}v_{1}\dots
v_{i-1}v_{i}v_{i+1}\dots v_{n}\nonumber\\
&=\alpha\;v_{0}v_{1}\dots v_{i-1}v_{i}v_{i+1}\dots v_{n}
\end{align}
Here we have used that $\alpha=\sum_{b\,n.n.\,i}[b]/[i-1]$ \cite{}, which gives us once again the value of the collapsed triangular cells, to wit:
\begin{equation}
\alpha=\sum_{b\,n.n.\,i}\frac{[b]}{[i-1]}=\frac{\abs{T_{i-1,b,i-1}}^{2}}{[i-1]^{2}}
\end{equation}
which validates the values for the collapsed cells we defined above.$\blacksquare$

We have yet to address the combination $C_{i}C^{\dag}_{j}$ for the case when the vertices upon which the operators act are separated by one position. This case is of particular interest since it is the first glimpse of the need to introduce the cup and cap operators. In this case we find:
 \begin{align}
 C_{i}C^{\dag}_{i+1}\eta
&=C_{i+1}C^{\dag}_{i}=\sum_{b}\frac{T_{i-1,i,b}\ov{T_{i,b,i+1}}}{\sqrt{[i][i+1][
i-1][b]}}\times\nonumber \\ &\times\;v_{0}v_{1}\dots v_{i-1}v_{b}v_{i+1}\dots v_{n}\neq\munite\\
 C_{i}C^{\dag}_{i-1}\eta
&=C_{i-1}C^{\dag}_{i}=\sum_{b}\frac{T_{b,i-1,i}\ov{T_{i-2,b,i-1}}}{\sqrt{[i-2][
i-1][b][i]}}\times\nonumber \\&\times\;v_{0}v_{1}\dots v_{i-2}v_{b}v_{i}\dots v_{n}\neq\munite
 \end{align}
 
This result conflicts with what we would desire if we are to find perfect analogy to the $SU(2)$ case, where this operation yields the identity. Careful inspection can provide us with the reason for this: the geometrical analogues of the creation and annihilation operators of $SU(2)$ are now the cup and cap operators of $SU(3)$, and the $SU(3)$ creation and annihilation operators behave differently since we are now on a two dimensional lattice built with triangular cells. A way to recover the result of $SU(2)$ requires that one considers the square of the combination of operators, that is to say:
\begin{widetext}
 \begin{align}
 C_{i}C^{\dag}_{i+1}C_{i}C^{\dag}_{i+1}\eta
&=\sum_{b,b'}\frac{T_{i-1,i,b}\ov{T_{i,b,i+1}}T_{i-1,b,b'}\ov{T_{b,b',i+1}}}{[
i-1][b][i+1]\sqrt{[i][b']}}\;v_{0}v_{1}\dots v_{i-1}v_{b'}v_{i+1}\dots v_{n}
\nonumber\\
 &=\sum_{b'}\frac{1}{\sqrt{[i][b']}}\frac{1}{[i-1][i+1]}\left([i-1][i][i+1]
\delta_{i,b'}+[i-1][i][b']\delta_{i-1,i+1}\right)\;v_{0}v_{1}\dots
v_{i-1}v_{b'}v_{i+1}\dots v_{n} \nonumber\\
 &=\eta + \sum_{b'}\frac{\sqrt{[i][b']}}{[i-1]}\delta_{i-1,i+1}\;v_{0}v_{1}\dots
v_{i-1}v_{b'}v_{i+1}\dots v_{n}\label{cuadratico1}
 \end{align}
 \begin{align}
 C_{i}C^{\dag}_{i-1}C_{i}C^{\dag}_{i-1}\eta
&=\sum_{b,b'}\frac{T_{i-2,i-1,b}\ov{T_{i,i-1,b}}T_{i,b',b}\ov{T_{i-2,b',b}}}{[
i-2][b][i]\sqrt{[i-1][b']}}\;v_{0}v_{1}\dots v_{i-2}v_{b'}v_{i}\dots v_{n}
\nonumber\\
 &=\sum_{b'}\frac{1}{\sqrt{[i][b']}}\frac{1}{[i-2][i]}\left([i-2][i][i-1]\delta_
{i-1,b'}+[i-2][i-1][b']\delta_{i-2,i}\right)\;v_{0}v_{1}\dots
v_{i-2}v_{b'}v_{i}\dots v_{n} \nonumber\\
 &=\eta + \sum_{b'}\frac{\sqrt{[i-1][b']}}{[i]}\delta_{i-2,i}\;v_{0}v_{1}\dots
v_{i-2}v_{b'}v_{i}\dots v_{n}\label{cuadratico2}
 \end{align}
 \end{widetext}
 In these calculations we have used the ``big pocket equation'' for the noncollapsed triangular cells we cited above and can also be understood as one of the three Kuperberg relations for $A_{2}$ tangles, in particular, the one involving the square tangle. In the context of paths one can clearly see what this operation entails: one of the paths created is the original path $\eta$ and the other one is a path that now has a back-and-forth backtracking in a similar fashion as it would be for paths in $SU(2)$. This makes clear what we said before: in $SU(3)$ one not only needs to introduce creation and annihilation operators for triangular sequences but one must retain, with minor modifications, analogues for the creation and annihilation operators for $SU(2)$ in order to properly study paths in the $SU(3)$ lattice.
 
 \subsection{Essential paths}
 As we mentioned in the introduction to this work, the objective of the path formalism is to provide a way of obtaining the bialgebra of a graph since their space of endomorphisms leads to one of the two basis of the vertical double triangle algebra \footnote{The space of endomorphisms of double triangles has at least two important bases: the vertical base, which is discussed here, and the horizontal base which naturally induces the coproduct structure of the bialgebra.} \cite{CoqueA2:2005,CoqueTrincheroDTE:2004}. The next key step in this process is finding the essential paths. 

 \begin{defi}\label{def:essential}
 An essential path $\xi$, of length $l$, is one for which {\bf both} of the following conditions are fulfilled:
 \begin{equation}
 C_{i}(\xi)=0,\quad\cup_{i}(\eta)=0,\qquad\mbox{both for all}\; i<n.
 \end{equation}
 \end{defi}
 This means that an essential path is one lacking not just in triangular sequences but also in back-and-forth like those in paths of $SU(2)$. Admissible triangles, i.e. the diagrammatical representation of the module $G\times\mathcal{A}(G)\rightarrow G$, are related to essential paths in the following way: the elements of $G$ are the start and end points of the path and the element of $\mathcal{A}(G)$ encodes the length of the path. The length of the path is encoded by the element of the $\mathcal{A}$ type graph since the triangular coordinates of the element tells one the amount of vertices of the path that should go in the direction of the arrows or opposite to them.
 
\unitlength 0.040cm 
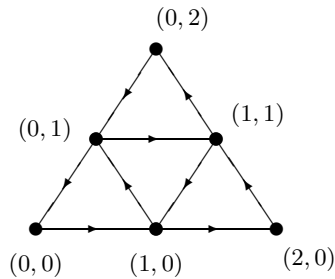
\begin{figure}[hhh]  
\begin{center}  
\begin{picture}(100,100)  
  
\put(-200,0){\begin{picture}(40,40)(-180,-15)  
\put(0,-12){\makebox(0,0){$(0,0)$}}  
\put(3,33){\makebox(0,0){$(0,1)$}}  
\put(0,0){\circle*{4}}  
\put(40,0){\circle*{4}}  
\put(20,30){\circle*{4}}  
\put(0,0){\vector(1,0){21}}  
\put(20,0){\line(1,0){20}}  
\put(40,0){\vector(-2,3){11}}  
\put(30,15){\line(-2,3){10}}  
\put(20,30){\vector(-2,-3){11}}  
\put(10,15){\line(-2,-3){10}}\end{picture}}

\put(-160,0){\begin{picture}(40,40)(-180,-15)  
\put(0,-12){\makebox(0,0){$(1,0)$}}  
\put(0,0){\circle*{4}}  
\put(40,0){\circle*{4}}  
\put(20,30){\circle*{4}}  
\put(0,0){\vector(1,0){21}}  
\put(20,0){\line(1,0){20}}  
\put(40,0){\vector(-2,3){11}}  
\put(30,15){\line(-2,3){10}}  
\put(20,30){\vector(-2,-3){11}}  
\put(10,15){\line(-2,-3){10}
\put(28,23){\makebox(0,0){$(1,1)$}}
\put(45,-25){\makebox(0,0){$(2,0)$}}}\end{picture}}  
  
\put(-180,30){\begin{picture}(40,40)(-180,-15)  
\put(0,0){\circle*{4}}  
\put(40,0){\circle*{4}}  
\put(20,30){\circle*{4}}  
\put(0,0){\vector(1,0){21}}  
\put(20,0){\line(1,0){20}}  
\put(40,0){\vector(-2,3){11}}  
\put(30,15){\line(-2,3){10}}  
\put(20,30){\vector(-2,-3){11}}  
\put(10,15){\line(-2,-3){10}
\put(23,25){\makebox(0,0){$(0,2)$}}}\end{picture}}  
  
\end{picture}  
\end{center}  
\caption{The $\mathcal{A}_{2}$ graph for $SU(3)_2$. The labeling used is that of the ``triangular coordinates'', which also count the number of $(\sigma,\ov{\sigma})$ generators involved in creating a path of a given length.}  
\label{A2SU3}  
\end{figure}

 \section{An additional formulation for the space of paths}
 
 As we have shown above, in order to have a complete description of the space of paths over an $SU(3)$ graph, one needs to expand the idea of a backtracking path in such a way so as to accept triangular and back-and-forth sequences. We see these sequences as ones intrinsic to the nature of $SU(3)$ graphs (for the triangular sequences) or as inheritance from the path formalism as applied to $SU(2)$ graphs (as is the case for back-and-forth sequences). We will show here that one could construct an alternative formulation for these operators that yields the very same essential paths while at the same time allowing for an interpretation that does not require grounding in $SU(2)$ type paths, this is to say, all operators (and paths generated by them) will be understood in terms of the fundamental triangles of $SU(3)$ graphs and only on them.
 
 Even though such an idea may at first seem trivial, it addresses an important subject regarding the calculation of essential paths. In previous works \cite{CoquereauxIsasiSchieber:2010,Evans:2009ar} the path formalism is introduced for $SU(3)$ and the many properties of the triangular cells are explained thoroughly. However, if one attempts to calculate the essential paths using only the operators described therein, one finds that for paths of any length that combines both generators $(\sigma, \ov{\sigma})$ it is impossible to find the essential paths without considering the cup and cap operators defined above and the collapsed triangular cells that they imply. This may lead one to ask whence are the collapsed cells in these works and if the properties studied in those works apply to the collapsed triangular cells. 
 
 \subsection{The trident and fork operators}
 
 The answer to both questions comes from some reflection on the idea of a backtracking path: due to the fact that its ADE graphs are bioriented, in $SU(2)$ a backtracking sequence can only be a back-and-forth sequence, but in $SU(3)$ we find that the idea must be expanded to include triangular sequences. However, since $SU(3)$ graphs are based in triangles \footnote{Here we are not considering the conjugate graphs, since for these graphs the cell system remains unknown.} it would be natural to expect that a backtracking be exchanged to a closed triangular sequence, that is to say, a round trip around a triangular cell. Replacing back-and-forth backtracks in this way allows us to replace the need for collapsed triangular cells for new operators to replace the cup and cap operators introduced above. These operators instead of adding and removing back-and-forth sequences will add and remove closed triangular ones. With this in mind we define:
 
 \begin{defi}
 The action of the trident and fork operators on a path $\eta$, is:
 \begin{equation}\label{eq:deftridente}
 \tridente_{i}(\eta)=\frac{T_{i-1,i,i+1}}{[i-1]}\delta_{i-1,i+2}\; v_{0}v_{1}\dots v_{i-1}v_{i+3}\dots v_{n},
 \end{equation}
  \begin{equation}\label{eq:deffork}
  \pitchfork_{i}(\eta)=\sum_{v_{b},v_{b'}}\frac{T_{i-1,b,b'}}{[i-1]}\; v_{0}v_{1}\dots v_{i-1}v_{b}v_{b'}v_{i-1}v_{i}\dots v_{n}.
  \end{equation}
  \end{defi}
  
  For the trident operator, $\tridente_{i}$ we have that it only gives a nonzero result if the sequence $v_{i-1}v_{i}v_{i+1}v_{i+2}$ is a closed triangle: the delta function enforces the equality of the first and last vertices of the sequence and the value of the triangular cell ensures that the vertices $v_{i-1}v_{i}v_{i+1}$ form a triangle. In any other case the operator annihilates the path. For the fork operator $\pitchfork_{i}$ the sums over the $v_{b}$ and $v_{b'}$ vertices are made so as to ensure that the sequences $v_{i-1}v_{b}v_{b'}v_{i-1}$ are closed triangles.

These operators can be seen to appear when one explores the Temperley-Lieb algebra associated to the creation and annihilation operators $C$ and $C^{\dag}$: with the Temperley-Lieb operator defined as $U_{i}=C^{\dag}_{i}C_{i}$, the analogous to the Yang-Baxter equation for $SU(3)$, $U_{i}U_{i+1}U_{i}-U_{i}=U_{i+1}U_{i}U_{i+1}-U_{i-1}=F_{i}$ leads to the definition of the $F_{i}$ operator:
\begin{align}
\pitchfork_{i}\tridente_{i}(\eta)=F_{i}(\eta)&=\sum_{b,b'}\frac{T_{i-1,i,i+1}\ov{T_{i-1,b,b'}}}{[i-1][i-1]}\times\nonumber\\&\times\;v_{0}v_{1}\dots v_{i-1}v_{b}v_{b'}v_{i-1}\dots v_{n}
\end{align}
This operator serves the same purpose as the cup-cap operator in the previous formulation, and just like the previous case, can be obtained by using the same Kuperberg relation as that case.

\begin{prop}
The trident and fork operators fulfill the following relations:
\begin{align}
 \tridente_{i\pm 1}\pitchfork_{i}= \tridente_{i} \pitchfork_{i\pm 1}=\beta\munite,\quad &\mbox{for}\; i\leq n\\
 \tridente_{i}\pitchfork_{i}=\alpha\beta\munite_{n},\quad &\mbox{for}\; i\leq n
\end{align}
\end{prop}

The proof of these relations is straightforward and follows from the definitions:
\begin{align}
 \tridente_{i+1}\pitchfork_{i}&= \sum_{b,b'}\frac{T_{b,b',i-1}\ov{T_{i-1,b,b'}}}{[i-1][b]}\delta_{b,i}\;v_{0}v_{1}\dots v_{i-1}v_{b}v_{i+1}\dots v_{n}\nonumber\\
 &= \sum_{b'}\frac{T_{i,b',i-1}\ov{T_{i-1,i,b'}}}{[i-1][i]}\;v_{0}v_{1}\dots v_{i-1}v_{i}v_{i+1}\dots v_{n}\nonumber\\
 &= \sum_{b'}\frac{\abs{T_{i-1,i,b'}}^{2}}{[i-1][i]}\;v_{0}v_{1}\dots v_{i-1}v_{i}v_{i+1}\dots v_{n}\nonumber\\
 &= \beta\eta\nonumber\\
 \tridente_{i-1}\pitchfork_{i}&= \sum_{b,b'}\frac{T_{i-1,i,b}\ov{T_{i,b,b'}}}{[i-1][i]}\delta_{i-1,b'}\;v_{0}v_{1}\dots v_{i-1}v_{i}v_{i+1}\dots v_{n}\nonumber\\
   &= \sum_{b}\frac{\abs{T_{i-1,i,b}}^{2}}{[i-1][i]}\;v_{0}v_{1}\dots v_{i-1}v_{i}v_{i+1}\dots v_{n}\nonumber\\
    &= \beta\eta
\end{align}
This relation is analogous to the topological simplifications that we found for the cup and cap operators, $\cup_{i\pm 1}\cap_{i}=\munite$. The other equality is related to the loop Kuperberg relation as we will show:
\begin{align}
\tridente_{i}\pitchfork_{i}&=\sum_{b,b'}\frac{T_{i-1,b,b'}\ov{T_{i-1,b,b'}}}{[i-1][i-1]}\;v_{0}v_{1}\dots v_{i-1}v_{i}v_{i+1}\dots v_{n}\nonumber\\
&= \sum_{b,b'}\frac{\abs{T_{i-1,b,b'}}^{2}}{[i-1][i-1]}\;v_{0}v_{1}\dots v_{i-1}v_{i}v_{i+1}\dots v_{n}\nonumber\\
&= \sum_{b}\frac{\beta[i-1][b]}{[i-1][i-1]}\;v_{0}v_{1}\dots v_{i-1}v_{i}v_{i+1}\dots v_{n}\nonumber\\
&=\beta\sum_{b}\frac{[b]}{[i-1]}\;v_{0}v_{1}\dots v_{i-1}v_{i}v_{i+1}\dots v_{n}\nonumber\\
&=\alpha\beta\eta.&\blacksquare
\end{align}

\subsection{Equivalence of the formulations}
Now that we have both sets of operators, we must show that they are interchangeable, that is, that one set can be written in terms of the other and that they both give the desired set of essential paths.

\begin{prop}
Given the two sets of operators $\{C,C^{\dag},\cup,\cap\}$ and $\{C,C^{\dag},\tridente,\pitchfork\}$ defined above, the following relations are satisfied for a path $\eta$:
\begin{align}
\tridente_{i}&=\cup_{i}C_{i}=\cup_{i}C_{i+1}\label{rel:trid-cup}\\
\pitchfork_{i}&=C^{\dag}_{i}\cap_{i}=C^{\dag}_{i+1}\cap_{i}\\
\cup_{i}&=\frac{1}{\beta}\tridente_{i}C^{\dag}_{i}=\frac{1}{\beta}\tridente_{i}C^{\dag}_{i+1}\label{rel:cup-trid}\\
\cap_{i}&=\frac{1}{\beta}C_{i}\pitchfork_{i}=\frac{1}{\beta}C_{i+1}\pitchfork_{i}
\end{align}
\end{prop}

Note that the definitions of the trident and fork operators will require the use of the collapsed triangular cells:

\paragraph{The definition of the trident operator in terms of the annihilation and cup operators:}
\begin{align}
\cup_{i}C_{i}(\eta)&=\frac{T_{i-1,i+1,i+2}T_{i-1,i,i+1}}{\sqrt{[i-1][i+2][i+1][i-1]}}\times\nonumber\\&\times\delta_{i-1,i+2}\;v_{0}v_{1}\dots v_{i-1}v_{i+3}\dots v_{n}\nonumber\\
&=\frac{T_{i-1,i+1,i-1}T_{i-1,i,i+1}}{\sqrt{[i-1][i-1][i+1][i-1]}}\times\nonumber\\&\times\;v_{0}v_{1}\dots v_{i-1}v_{i+3}\dots v_{n}\nonumber\\
&=\frac{T_{i-1,i,i+1}}{[i-1]}\;v_{0}v_{1}\dots v_{i-1}v_{i+3}\dots v_{n}=\tridente_{i}(\eta)\\
\cup_{i}C_{i+1}(\eta)&=\frac{T_{i-1,i,i+2}T_{i,i+1,i+2}}{\sqrt{[i-1][i+2][i][i+2]}}\times\nonumber\\&\times\delta_{i-1,i+2}\;v_{0}v_{1}\dots v_{i-1}v_{i+3}\dots v_{n}\nonumber\\
&=\frac{T_{i-1,i,i-1}T_{i,i+1,i-1}}{\sqrt{[i-1][i-1][i][i-1]}}\;v_{0}v_{1}\dots v_{i-1}v_{i+3}\dots v_{n}\nonumber\\
&=\frac{T_{i-1,i,i+1}}{[i-1]}\;v_{0}v_{1}\dots v_{i-1}v_{i+3}\dots v_{n}=\tridente_{i}(\eta).
\end{align}

\paragraph{The definition of the fork operator in terms of the creation and cap operators:}
\begin{align}
C^{\dag}_{i}\cap_{i}(\eta)&=\sum_{b,b'}\frac{\ov{T_{i-1,b,i-1}}}{[i-1]}\frac{\ov{T_{i-1,b',b}}}{\sqrt{[i-1][b]}}\times\nonumber\\&\times\; v_{0}v_{1}\dots v_{i-1}v_{b'}v_{b}v_{i-1}v_{i}\dots v_{n} \nonumber\\
&=\sum_{b,b'}\frac{\ov{T_{i-1,b',b}}}{[i-1][b]}\; v_{0}v_{1}\dots v_{i-1}v_{b'}v_{b}v_{i-1}v_{i}\dots v_{n}\\&=\pitchfork_{i}(\eta)\nonumber\\
C^{\dag}_{i+1}\cap_{i}(\eta)&=\sum_{b,b'}\frac{\ov{T_{i-1,b,i-1}}}{[i-1]}\frac{\ov{T_{i-1,b',b}}}{\sqrt{[i-1][b]}}\times\nonumber\\&\times\; v_{0}v_{1}\dots v_{i-1}v_{b'}v_{b}v_{i-1}v_{i}\dots v_{n} \nonumber\\
&=\sum_{b,b'}\frac{\ov{T_{i-1,b',b}}}{[i-1][b]}\; v_{0}v_{1}\dots v_{i-1}v_{b'}v_{b}v_{i-1}v_{i}\dots v_{n}\\&=\pitchfork_{i}(\eta).\nonumber
\end{align}

For the definition of the cup and cap operators we use equation \ref{eq:smallpock}. 

\paragraph{The definition of the cup operator in terms of the creation and trident operators:}
\begin{align}
\tridente_{i}C^{\dag}_{i}(\eta)&=\sum_{b}\frac{\ov{T_{i-1,b,i}}}{\sqrt{[i-1][i]}}\frac{T_{i-1,b.i}}{\sqrt{[i-1][i-1]}}\times\nonumber\\&\times\;\delta_{i-1,i+1}v_{0}v_{1}\dots v_{i-1}v_{i+2}\dots v_{n}\nonumber\\
&=\sum_{b}\frac{\abs{T_{i-1,b,i}}^{2}}{[i-1]\sqrt{[i-1][i]}}\;\delta_{i-1,i+1}\times\nonumber\\&\times\;v_{0}v_{1}\dots v_{i-1}v_{i+2}\dots v_{n}\nonumber\\
&=\beta\frac{T_{i-1,i,i+1}}{[i-1]}\delta_{i-1,i+1}\;v_{0}v_{1}\dots v_{i-1}v_{i+2}\dots v_{n}\nonumber\\&=\beta\cup_{i}(\eta)
\end{align}
\begin{align}
\tridente_{i}C^{\dag}_{i+1}(\eta)&=\sum_{b}\frac{\ov{T_{i,b,i+1}}}{\sqrt{[i][i+1]}}\frac{T_{i-1,b.i}}{\sqrt{[i-1][i-1]}}\times\nonumber\\&\times\;\delta_{i-1,i+1}v_{0}v_{1}\dots v_{i-1}v_{i+2}\dots v_{n}\nonumber\\
&=\sum_{b}\frac{\abs{T_{i-1,b,i}}^{2}}{[i-1]\sqrt{[i-1][i]}}\;\delta_{i-1,i+1}\times\nonumber\\&\times\;v_{0}v_{1}\dots v_{i-1}v_{i+2}\dots v_{n}\nonumber\\
&=\beta\frac{T_{i-1,i,i+1}}{[i-1]}\delta_{i-1,i+1}\;v_{0}v_{1}\dots v_{i-1}v_{i+2}\dots v_{n}\nonumber\\&=\beta\cup_{i}(\eta)
\end{align}

\paragraph{The definition of the cap operator in terms of the annihilation and fork operators:}
\begin{align}
C_{i}\pitchfork_{i}&=\sum_{b,b'}\frac{\ov{T_{i-1,b,b'}}}{[i-1]}\frac{T_{i-1,b,b'}}{\sqrt{[i-1][b']}}\times\nonumber\\&\times\;v_{0}v_{1}\dots v_{i-1}v_{b}v_{b'}v_{i-1}v_{i}\dots v_{n}\nonumber\\
&=\beta\sum_{b'}\frac{\ov{T_{i-1,b,i-1}}}{[i-1]}\;v_{0}v_{1}\dots v_{i-1}v_{b'}v_{i-1}v_{i}\dots v_{n}\nonumber\\
&=\beta\cap_{i}(\eta)
\end{align}
\begin{align}
C_{i+1}\pitchfork_{i}&=\sum_{b,b'}\frac{\ov{T_{i-1,b,b'}}}{[i-1]}\frac{T_{b,b',i-1}}{\sqrt{[i-1][b]}}\times\nonumber\\&\times\;v_{0}v_{1}\dots v_{i-1}v_{b}v_{b'}v_{i-1}v_{i}\dots v_{n}\nonumber\\
&=\beta\sum_{b'}\frac{\ov{T_{i-1,b,i-1}}}{[i-1]}\;v_{0}v_{1}\dots v_{i-1}v_{b}v_{i-1}v_{i}\dots v_{n}\nonumber\\
&=\beta\cap_{i}(\eta).&\blacksquare
\end{align}

\subsubsection{The space of essential paths}
The remaining property that we must require for both formulations to be equivalent is for the space of essential paths to be the same. 

\begin{prop}
The kernels of both sets of operators: $\{C,C^{\dag},\cup,\cap\}$ and $\{C,C^{\dag},\tridente,\pitchfork\}$ are compatible in the sense that the essential paths of each formulation are associated to the same set of admissible triangles. 

A path $\eta$ is essential for the set of operators $\{C,C^{\dag},\tridente,\pitchfork\}$ if:
 \begin{equation}
 C_{i}(\eta)=0,\quad\tridente_{i}(\eta)=0,\qquad\mbox{both for all}\; i<n.
 \end{equation}
\end{prop}
Paths of length $l=0,1$ are trivially essential on both sets of operators since the action of all annihilation operators (be them $C$, cups or tridents) yields zero result by definition. Also, since both sets of operators include the creation and annihilation operators $C_{i}, C^{\dag}_{i}$ then the paths of interest are those in the kernel of the cup and cap or trident and fork operators.

Let $\eta$ be a path in $\{\ker(\cup_{i})\}\cap\{\ker(C_{i})\}$, then equation \ref{rel:trid-cup} immediately ensures that $\eta\in\{\ker(\tridente_{i})\}$.

Now let $\eta\in\{\ker(\tridente_{i})\}\cap\{\ker(C_{i})\}$ this means that $\eta=\sum_{j}\alpha_{j} (v_{0}v_{1}\dots v_{i-1}v_{i}v_{i+1}v_{i-1}\dots v_{n})$ with 
\begin{align}
\tridente_{i}(\eta)&=\sum_{j}\alpha_{j}\left(\frac{T_{i-1,i,i+1}}{[i-1]}\times\right)_{j}\nonumber\\& \times \;v_{0}v_{1}\dots v_{i-1}v_{i+3}\dots v_{n}=0
\end{align}
which is true $\forall\tridente_{i}$. Since this path has two triangular sequences $v_{i-1}v_{i}v_{i+1}$ and $v_{i}v_{i+1}v_{i-1}$ it does not immediately belong in $\{\ker(\cup_{i})\}\cap\{\ker(C_{i})\}$. We will now then act on it with $C_{i}$ to eliminate at least one of those triangular sequences thus creating a new path $\eta'$. Doing so we get:
\begin{align}
\eta'=C_{i}(\eta)&=\sum_{j}\alpha_{j}\left(\frac{T_{i-1,i,i+1}}{\sqrt{[i-1][i+1]}}\times\right.\nonumber\\&\left.\times \;v_{0}v_{1}\dots v_{i-1}v_{i+1}v_{i-1}\dots v_{n}\right)_{j}
\end{align}
the paths we have obtained now are indeed in $\{\ker(C_{i})\}$ since they only have a back-and-forth sequence. Acting upon them with the cap operator we get:
\begin{align}
\cup_{i}C_{i}(\eta)&=\sum_{j}\alpha_{j}\left(\frac{T_{i-1,i,i+1}}{\sqrt{[i-1][i+1]}}\frac{T_{i-1,i+1,i-1}}{[i+1]}\times\right)_{j}\nonumber\\&\times \;v_{0}v_{1}\dots v_{i-1}v_{i+3}\dots v_{n}\nonumber\\
&=\sum_{j}\alpha_{j}\left(\frac{T_{i-1,i,i+1}}{[i+1]}\times\right)_{j}\nonumber\\&\times \;v_{0}v_{1}\dots v_{i-1}v_{i+3}\dots v_{n}
\end{align}
which is zero since the values of the $\alpha_{j}$ coefficients ensure cancellation by hypothesis. 

So far we have shown the path $C_{i}(\eta)$ is in $\{\ker(\cup_{i})\}\cap\{\ker(C_{i})\}$, and therefore essential. This new path starts at $v_0$ and ends $v_n$ just like $\eta$, but two of its oriented edges have been replaced by one edge with the opposite orientation, this substitution is naturally induced in the structure by the fundamental intertwiners $\sigma\otimes\sigma\rightarrow\ov{\sigma}$ and $\ov{\sigma}\otimes\ov{\sigma}\rightarrow\sigma$. There is one linear combination of elementary paths of length $(\alpha+2,\beta-1)$ or $(\alpha-1,\beta+2)$ for each linear combination of elementary paths of length $(\alpha,\beta)$ that will be essential, and even more the set of coefficients is the same. 

In light of this we can see that both sets of paths are equivalent, in the sense that both paths $\eta$ and $\eta'$ are associated to the same admissible triangle, with the only difference being that one interprets the value of the $\mathcal{A}(G)$ element (that is related to the length of the path) in different fashions: in one we take combinations of generators $\sigma\ov{\sigma}$ or $\ov{\sigma}\sigma$ as meaning sequences of length $(1,1)$ and in the other one we use the intertwines to change one of the generators into two of the other kind, changing $(1,1)\rightarrow(3,0)$ or $(1,1)\rightarrow(0,3)$. Since they are related in this way to the same admissible triangle we can ensure that the space generated by them is the same.$\blacksquare$

\section{Discussion}

As was shown in Trinchero's work for $SU(2)$ \cite{Trinchero:2010yr}, a key concept in the path formalism for the calculation of the bialgebra associated with an $ADE$ type graph is the idea of backtracking sequences, since it is their existence (or absence) what determines whether a path is essential and thus whether it translates an admissible triangle of the fusion algebra. Subsequent works \cite{CoquereauxIsasiSchieber:2010} \cite{Evans:2009ar,Evans:2009ud} have shown that in the case of $SU(3)$ it is necessary to extend the idea of a backtracking sequence to consider open triangular sequences and they introduce an annihilation operator that involves what we have called open triangular sequences. This new behavior of backtrackings in $SU(3)$ is due to the fact that the graph is embedded in a two dimensional triangular lattice.

Here we have shown that for $SU(3)$ simply laced $ADE$ graphs it is possible to choose between two different but equivalent types of fundamental backtracking paths. One can work, either considering open triangular sequences (as done in previous works) and also closed triangular sequences as fundamental backtracking sections or replacing the latter with back-and-forth backtracks not unlike those considered for $SU(2)$ graphs.  We further show that once the selection is made one can and should write the other type of sequence in terms of the one selected, this is to say that if, for example, one chooses to consider open and closed triangles as fundamental then one must consider back-and-forth backtracks as ones resulting from creating a closed triangular sequence and then removing the appropriate open triangle. 

These two options for the fundamental backtracking section generate a pair of equivalent (in the sense that we have explained above) formulations for essential paths on an $SU(3)$ simply laced graph. These formulations are encoded by the two equivalent sets of operators acting on the space of paths: the creation and annihilation operators acting on open triangular sequences plus the cup and cap operators acting on back-and-forth sequences or, equivalently, the creation and annihilation operators plus the fork and the trident acting on closed triangular sequences.  The equivalence is established as follows: an essential path is given by a admissible triangle $(a,n,b)$, meaning that there is a path (in general a linear combination of elementary paths) starting from vertex $a$ and ending on vertex $b$ of the graph. The $n$ term of the triad is an element of the $A$-type graph with the same generalized Coxeter number and can be understood to encode the length of the path in at least two different ways: either by the sum of the triangular coordinates $(\alpha,\beta)$ of the $n$ vertex of the graph (as displayed in Figure \ref{A2SU3}) or by providing a sort of inventory of the number of edges of one type (e.g. $\ov{\sigma}$) that must be exchanged into the other type ($\sigma$) in order to restrict ones measure of length to edges of a single graph.

One could be tempted to assign more significance to one formulation over the other: it should be clear that the formulation that employs the set of operators $\{C,C^{\dag},\cup,\cap\}$ (and thus the first definition of length for the element $n$ of the admissible triangle) is one that, at least intuitively yields path of minimal length connecting two vertices of the graph. However one must consider that the intuitive definition of length of simply counting the number of edges is not the only possible one, and as we mentioned one can take into account the type of edge that is involved (either going with or against the arrows of the graph).

Counting the number of edges in each orientation is related to the number of generators $\sigma$ or $\bar{\sigma}$ that are required to construct the path and provides us with a clearer understanding of the possibilities allotted to us in the selection of one formulation over the other. Through the use of the relations of the generators $\sigma\times\sigma\rightarrow\ov{\sigma}$ (and its conjugate), one can see that if one is given a path of length $l=\alpha+\beta$, with $\alpha,\beta$ counting the number of vertices in each direction of the arrows, it is possible to find an equivalent path connecting the first and last vertices of the path with edges of only one type \footnote{This relation of the generators is related to the fusion algebra and is the basis of the creation and annihilation operators $C$ and $C^{\dag}$ since one can check that they exchange one edge of one orientation for two of the other and viceversa, respectively.}.  

This reasoning also brings to light the connection between closed triangular sequences and back-and-forth ones that lies at the center of this work. If we consider the morphism of generators $\sigma\times\ov{\sigma}$ that is associated to a back-and-forth sequence of vertices we can find $\sigma\times\ov{\sigma}\rightarrow(\ov{\sigma}  \times\ov{\sigma})\times\ov{\sigma}=\ov{\sigma}\times(\ov{\sigma}\times\ov{\sigma})\rightarrow\ov{\sigma}\times\sigma\rightarrow(\sigma\times\sigma)\times\sigma=\sigma\times  (\sigma\times\sigma)=\sigma\times\ov{\sigma}$. This shows us that a back-and-forth sequence is related to either a trivial identity operation, to a cup-cap operation (as we already showed in equations \ref{cuadratico1} and \ref{cuadratico2}) or to a closed triangular sequence of edges all of the same generator. It is this relation that allows for the equivalence of formulations and while it forces one to be careful with ones definition of the length of an essential path it allows for a definition of backtracking sequences that can be seen as more natural for $SU(3)$ graphs and does requires no knowledge of the treatment for $SU(2)$ graphs. This peculiar behavior for $SU(3)$ is a consequence of the nature of its fundamental intertwiners, as has been already shown in \cite{CoquerauxSchieber:2009}, since they are lead to the definitions of the coherence equations that allow the construction of the cell system of the graph (i.e. the big pocket and small pocket equations discussed in \cite{CoquereauxIsasiSchieber:2010}.).

\bibliographystyle{unsrt}
\bibliography{bibliography-paths-su3} 

\begin{thebibliography}{10}

\bibitem{Ocneanu:paths}
Adrian Ocneanu.
\newblock Paths on coxeter diagrams: fron platonic solids and singularities to
  minimal models and subfactors.
\newblock In {\em Lectures on Operator Theory}, volume~33 of {\em Fields
  Institute Monographs}, pages 245--323. American Mathematical Society, 1999.

\bibitem{Ocneanu:2000}
Adrian Ocneanu.
\newblock The classification of subgroups of quantum {${\rm SU}(N)$}.
\newblock In {\em Quantum symmetries in theoretical physics and mathematics
  (Bariloche, 2000)}, volume 294 of {\em Contemp. Math.}, pages 133--159. Amer.
  Math. Soc., Providence, RI, 2002.

\bibitem{CoqueGarciaTrinchero:1999}
R.~Coquereaux, A.~O. Garcia, and R.~Trinchero.
\newblock Hopf stars, twisted hopf stars and scalar products on quantum spaces.
\newblock {\em J. Geom. Phys.}, 36:22--59, 2000.

\bibitem{CoqueTrincheroDTE:2004}
Robert Coquereaux and Roberto Trinchero.
\newblock On quantum symmetries of ade graphs.
\newblock {\em Adv. Theor. Math. Phys.}, 8:189--216, 2004.

\bibitem{EvansBockenhauerIII:1998}
Jens Bockenhauer and David~E. Evans.
\newblock Modular invariants, graphs and alpha-induction for nets of
  subfactors. iii.
\newblock {\em Commun. Math. Phys.}, 205:183--228, 1999.

\bibitem{EvansPintoSubfqctor:2003}
D.~E. Evans and P.~R. Pinto.
\newblock Subfactor realisation of modular invariants.
\newblock {\em Commun. Math. Phys.}, 237:309--363, 2003.

\bibitem{Coque6jsymbols:2006}
Robert Coquereaux.
\newblock Racah - wigner quantum 6j symbols, ocneanu cells for a(n) diagrams,
  and quantum groupoids.
\newblock 2005.

\bibitem{CoquereauxIsasiSchieber:2010}
R.~{Coquereaux}, E.~{Isasi}, and G.~{Schieber}.
\newblock {Notes on TQFT Wire Models and Coherence Equations for SU(3)
  Triangular Cells}.
\newblock {\em SIGMA}, 6:99, December 2010.

\bibitem{Evans:2009ar}
D.~E. {Evans} and M.~{Pugh}.
\newblock {Ocneanu Cells and Boltzmann Weights for the SU(3) ADE Graphs}.
\newblock {\em Munster J. Math. 2 (2009), 95-142}, 2:95--142, June 2009.

\bibitem{CaicedoIsasiPineda:2015}
M.~Caicedo, E.~Isasi, and J.~A. Pineda.
\newblock {To be published}.
\newblock 2015.

\bibitem{Note1}
We remind the reader that each admissible triangle obtained from the fusion
  algebra is in one to one correspondence to an essential path.

\bibitem{Note2}
Note that this label can be interpreted as the length of the graph in different
  ways. This is due to the fact that the path lies in the lattice of the
  $SU(3)$ graph.

\bibitem{Trinchero:2010yr}
R.~Trinchero.
\newblock {Paths on graphs and associated quantum groupoids*}.
\newblock {\em Revista de la Unión Matemática Argentina}, 51:147--170, 2010.

\bibitem{kuperberg1996}
Greg Kuperberg.
\newblock Spiders for rank $2$ lie algebras.
\newblock {\em Communications in Mathematical Physics}, 180(1):109--151, 1996.

\bibitem{kauffman1991knots}
L.H. Kauffman.
\newblock {\em Knots and Physics}.
\newblock K \& E series on knots and everything. World Scientific, 1991.

\bibitem{Note3}
The space of endomorphisms of double triangles has at least two important
  bases: the vertical base, which is discussed here, and the horizontal base
  which naturally induces the coproduct structure of the bialgebra.

\bibitem{CoqueA2:2005}
R.~Coquereaux.
\newblock The a2 ocneanu quantum groupoid.
\newblock {\em Contemporary Mathematics}, 376, 2005.

\bibitem{Note4}
Here we are not considering the conjugate graphs, since for these graphs the
  cell system remains unknown.

\bibitem{Evans:2009ud}
David~E. Evans and Mathew Pugh.
\newblock {SU(3)-Goodman-de la Harpe-Jones subfactors and the realisation of
  SU(3) modular invariants}.
\newblock {\em Rev.Math.Phys.}, 21:877--928, 2009.

\bibitem{Note5}
This relation of the generators is related to the fusion algebra and is the
  basis of the creation and annihilation operators $C$ and $C^{\protect \dag }$
  since one can check that they exchange one edge of one orientation for two of
  the other and viceversa, respectively.

\bibitem{CoquerauxSchieber:2009}
R.~{Coquereaux} and G.~{Schieber}.
\newblock {Quantum Symmetries for Exceptional SU(4) Modular Invariants
  Associated with Conformal Embeddings}.
\newblock {\em SIGMA}, 5:44, April 2009.

\end{thebibliography}
\end{document}